\documentclass{aa}

\title{On marginally resolved objects in optical interferometry}
\author{R\'egis Lachaume}
\institute{Laboratoire d'Astrophysique de Grenoble}
\date{Received 13 November 2002 / Accepted 19 December 2002}
\authorrunning{Lachaume, R.}
\titlerunning{On marginally resolved objects in optical interferometry}
\offprints{%
   R\'egis Lachaume\\
   \email{Regis.Lachaume@obs.ujf-grenoble.fr}%
}

\def\correct#1{#1}

\abstract{%
   With the present and soon-to-be breakthrough of optical interferometry,
   countless objects shall be within reach of interferometers; yet, most of
   them are expected to remain only marginally resolved with hectometric
   baselines.

   In this paper, we tackle the problem of deriving the properties of a
   marginally resolved object from its optical visibilities.  We show that they
   depend on the moments of flux distribution of the object: centre, mean
   angular size, asymmetry, and curtosis.  We also point out that the visibility
   amplitude is a second-order phenomenon, whereas the phase is a combination
   of a first-order term, giving the location of the photocentre, and a
   third-order term, more difficult to detect than the visibility amplitude,
   giving an asymmetry coefficient of the object.  We then demonstrate that
   optical visibilities are not a good model constraint while the object stays
   marginally resolved, unless observations are carried out at different
   wavelengths.  Finally, we show an application of this formalism to
   circumstellar discs.

}

\usepackage[fleqn]{amsmath}
\usepackage{amssymb,bm}

\usepackage{graphicx}
\graphicspath{{./M3304fig/}}

\usepackage{natbib}
\newcommand\diff[2][]{\ensuremath{\mathrm{d}^{#1}#2}}
\newcommand\idiff[2][]{\ensuremath{\,\diff[#1]#2}}

\def\vec#1{\bm{#1}}
\def\valpha{\vec{\alpha}}
\def\valphar{\vec{\alpha}-\vec{\alpha}_0}
\def\sp{\!\cdot\!}
\def\vuvalpha{\vu\sp\valpha}
\def\vu{\vec{u}}
\def\vB{\vec{B}}
\def\equiv#1{\mathcal{O}(#1)}
\DeclareMathOperator{\real}{Re}
\DeclareMathOperator{\imag}{Im}

\def\vufirst{\vu_1}
\def\vulast{\vu_n}
\def\vuvalphafirst{\valpha\sp\vufirst}
\def\vuvalphalast{\valpha\sp\vulast}
\def\vuvalpharfirst{(\valphar)\sp\vufirst}
\def\vuvalpharlast{(\valphar)\sp\vulast}

\def\Mom#1{\vec{M}_#1}
\def\Mr#1{\vec{M'}_#1}
\def\vi{\ensuremath{\vec{i}}}

\def\uone{\ensuremath{\vec{u}_{12}}}
\def\utwo{\ensuremath{\vec{u}_{23}}}
\def\uthree{\ensuremath{\vec{u}_{31}}}
\def\clph{\ensuremath{\bar{\phi}}}

\def\Nlambda{\ensuremath{N_\lambda}}
\def\Np{\ensuremath{N_\mathrm{p}}}

\def\Vstar{\ensuremath{V_\star}}
\def\Fstar{\ensuremath{F_\star}}

\def\Vtherm{\ensuremath{V_\mathrm{th}}}
\def\Ftherm{\ensuremath{F_\mathrm{th}}}
\def\Stherm{\ensuremath{S_\mathrm{th}}}

\def\Vscat{\ensuremath{V_\mathrm{s}}}
\def\Fscat{\ensuremath{F_\mathrm{s}}}

\def\Vtot{\ensuremath{V_\mathrm{tot}}}
\def\Ftot{\ensuremath{F_\mathrm{tot}}}

\def\Dtherm{\ensuremath{D_\mathrm{th}}}
\def\Dstar{\ensuremath{D_\star}}
\def\qapp{\ensuremath{q_\mathrm{app}}}
\def\M#1#2{\ensuremath{{M'}_#1^#2}}

\def\Mdot{\ensuremath{\dot M}}
\def\Mstar{\ensuremath{M_*}}
\def\AU{\ensuremath{\mathrm{AU}}}
\def\Msun{\ensuremath{M_\odot}}
\def\Rsun{\ensuremath{R_\odot}}

\def\yr{\ensuremath{\mathrm{yr}}}

\begin{document}

\maketitle


\section{Introduction}

Optical systems are fundamentally limited in angular resolution by their
spatial extent because of diffraction.  Very soon, the idea of combining light
coming from two ``distant'' telescopes was investigated in order to overcome the
limitation in size of single pupils.
\citet{Michelson91,Michelson20,Michelson21} derived the angular diameters of
some solar-system bodies and stars by measuring the contrast of the fringes
(called visibility amplitude) obtained when interfering light comes from two
apertures: \correct{this contrast is maximum when these apertures are closest
and decreases with the distance between telescopes (called the baseline).  The
baseline at which the fringes disappear holds information on the angular extent
of the source.} After some time,  the mid-seventies saw the come-back of
optical interferometry with \citet{Hanbury74,Labeyrie75}, but it long stayed
confined to bright and simple objects, mostly stellar diameters and multiple
systems.  It is all the more frustrating as the theory of interferometry allows
image reconstruction and as radio arrays achieved this goal within a few
decades of existence: the atmospheric turbulence and the nature of light both
lead to complex optical designs and have slowed the development of optical
interferometry.  Moreover, the shift of the fringes (called phase) is
completely \correct{blurred} by the atmosphere, so techniques to retrieve phase
information ---necessary for imaging capacities--- needed additional
investigation.

Recently, the Palomar Testbed Interferometer \citep[PTI,][]{Colavita99} and the
Infrared Optical Telescope Array \citep[IOTA,][]{Carleton94} allowed us to probe
circumstellar matter in star-forming regions
\citep{Malbet98,Akeson00,Malbet01p,Akeson02}, giving some constraints on the
geometry of these objects.  With the advent of the Very Large Telescope
Interferometre \citep[VLTI,][]{Glindemann00} and the \citep[KI,][]{Colavita01}
we are expecting a much higher accuracy with their large pupils (8--10\,m), and
good constraints on objects thanks to the number of baselines available
and partial phase information; yet they will not allow direct image
reconstruction very soon because recombination will be first performed with two
or three telescopes.  The Navy Prototype Optical Interferometre
\citep[NPOI,][]{Armstrong98}, the CHARA array \citep{tenBrummelaar00p},
\correct{and the Cambridge Optical Aperture Synthesis Interferometer}
\citep[COAST,][]{Haniff00p} provide imaging capacities with a
\correct{multi-telescope} recombination, but with a lower sensitivity that
renders faint object science difficult.  

We are clearly entering a phase in which more than an apparent diameter is
measured but no imaging is performed; in this context, observers and modellers
use interferometric observations as a constraint on models
\citep[e.g.]{Malbet01p,Akeson02,Lachaume02},  but their link with the geometry
of the object remains unclear; it is still quite common to think in terms of
diameter.  For instance, \citet{Monnier02} \correct{link the uniform disc
equivalent diameter derived from the IR interferometric observations of young
stellar objects with the physical radius of their supposed inner hole.}
\correct{The phase also raises problems of geometrical interpretation.  Since
it is blurred by the atmosphere,  one uses the closure phase to retrieve
partial information on it\footnote{Other techniques are also
the use of a close object as a phase reference or the differential phase with
spectroscopy, but we shall not discuss them in this paper.}:  the principle is
to add the phases over a triplet of baselines provided by three telescopes,
which allows one to cancel atmospheric terms.  It is generally used either in image
reconstruction, at NPOI for instance, or as a model constraint.  Geometrically
speaking, it is a diagnosis of asymmetry, but its accurate meaning is seldom 
made clear enough.}

In this paper,  we connect the \correct{visibility amplitude and phase} with
the geometry of the object, \correct{which allows us to retrieve information in a
model-independent fashion.} In Sect.~\ref{sec:geom}, we establish a series
development of \correct{these quantities} involving the moments of the flux
distribution, the first ones being the location of the photocentre, the spatial
extent (diameter), and the asymmetry coefficient (skewness).  It appears as a
generalisation of the widespread diameter measurement.  We then draw the
consequences of the formalism in terms of observation and modelling.  In
Sect.~\ref{sec:disc} we apply this development to circumstellar discs with two
examples: the case of an object characterised by more than one diameter (star,
thermal flux and scattered light) and the measurement of the radial temperature
law in these discs.


\section{Visibility and object geometry}
\label{sec:geom}

The Zernicke-van~Cittert theorem links the complex visibility $V$ to the 
normalised flux distribution $I$ of the object:
\begin{equation}
   V(\vu) = \iint I(\alpha) \exp\left( -2\pi i \vuvalpha \right) \idiff[2]\valpha,
\end{equation}
where $\vu$ is linked to the projected baseline $\vB$ and the wavelength 
$\lambda$ by $\vu = \lambda^{-1} \vB$, and $\valpha$ the angular location on the 
sky.  Optical interferometry usually deals with the square amplitude $|V|^2$ 
and the phase $\phi$, given by
\begin{subequations}
\begin{align}
   |V|^2     &= (\real V)^2 + (\imag V)^2,\\
   \tan\phi  &= (\imag V)/(\real V).
\end{align}
\end{subequations}
\correct{In interferometry two extreme cases are usually dealt with:  on the one
hand, an object is fully resolved if its angular size is of the order of
$B/\lambda$ as it would be with a single dish telescope of size $B$.  In that
case, the visibility is arbitrary and highly depends on the shape of the
target.  On the other hand, a point-like source is not resolved and its
visibility is $V = 1$.  Between these two cases, the object is said to be marginally
resolved and $V$ is close to unity; its size is a fraction of $B/\lambda$.}  In
such a case, most of the flux is located in a zone where  $|\vuvalpha| \ll 1$,
so we carry out a series development to derive the real and imaginary parts of
$V$:
\begin{subequations}
\begin{align}
   \real V(u) &= \iint I(\valpha) \left( 1 - 2\pi^2 (\vuvalpha)^2 \right) \idiff[2]\valpha,\\
   \imag V(u) &= \iint I(\valpha) \left( 2\pi (\vuvalpha) - \frac43\pi^3 (\vuvalpha)^3 \right) \idiff[2]\valpha.
\end{align}
\end{subequations}
These expressions drive us to define the $n$-th moment of the flux distribution
as a symmetric tensor unambiguously defined by
\begin{equation}
   \Mom{n} \sp \vufirst \cdots \vulast = \iint I(\valpha) (\vuvalphafirst) \cdots (\vuvalphalast) \idiff[2]\valpha
\end{equation}
\correct{for any set of vectors $\vu_1$, $\cdots$, $\vu_n$. (The expressions
for the first two moments are given in a Cartesian frame in 
Appendix~}\ref{ap:fm}).  With such a definition, the real and imaginary parts 
of $V$ become
\begin{subequations}
\begin{align}
   \real V(u) &= 1 - 2\pi^2 \Mom{2} \sp\vu \sp\vu,\\
   \imag V(u) &= 2\pi \Mom{1} \sp\vu - \frac43\pi^3 \Mom{3}\sp\vu\sp\vu\sp\vu,
\end{align}
\end{subequations}
which, after a few calculations, give 
\begin{subequations}
\begin{gather}
   |V|^2 = 1 - 4\pi^2\left[ \Mom{2}\sp\vu \sp\vu - (\Mom{1}\sp\vu)^2 \right] \label{eq:V:1},\\
   \begin{split}
      \phi  &= -2\pi (\Mom{1}\sp\vu) + \frac43\pi^3 \left[ \Mom{3}\sp\vu\sp\vu\sp\vu \right.\\
            &\qquad \left. - 3(\Mom{1}\sp\vu)(\Mom{2}\sp\vu\sp\vu) + 2(\Mom{1}\sp\vu)^3\right].
   \end{split}
   \label{eq:phi:1}
\end{gather}
\end{subequations}

These formulae are quite annoying because the visibility amplitude apparently
depends on the location of the photocentre of the object, given by the first
moment $\Mom{1}$ (see Appendix~\ref{ap:fm}), and therefore on the pointing 
accuracy of the instrument.  In order to cancel this apparent dependence, we 
define the moments in respect to the photocentre of angular location 
$\valpha_0$ by
\begin{equation}
\begin{split}
   \Mr{n}\sp\vufirst\cdots\vulast = &\iint I(\valpha) \vuvalpharfirst\\ 
                                    &\qquad\cdots \vuvalpharlast\idiff[2]\valpha
\end{split}
\end{equation}
and perform a frame change on Eqs.~(\ref{eq:V:1},\ref{eq:phi:1}) to obtain
\begin{subequations}
\begin{align}
   |V|^2 &= 1 - 4\pi^2 \left( \Mr{2}\sp\vu\sp\vu \right) + \equiv{u^4}, \label{eq:V:2}\\
   \phi  &= -2\pi \left( \Mom{1}\sp\vu \right) + \frac43\pi^3 \left( \Mr{3}\sp\vu\sp\vu\sp\vu \right) + \equiv{u^5}.\label{eq:phi:2}
\end{align}
\end{subequations}
In some applications,  a higher-order development is needed;  it is given in 
Appendix~\ref{sec:hod}.

\correct{Since the phase is not directly measured because of atmospheric
turbulence, the closure phase is used instead.  With three telescopes
labelled 1, 2, and 3 simultaneously providing the baselines $\uone$, 
$\utwo$, and $\uthree$, the closure phase reads}
\begin{equation}
   \clph = \phi(\uone) + \phi(\utwo) + \phi(\uthree).
\end{equation}
Applying Eq.~\ref{eq:phi:2}, we derive a concise expression:
\begin{equation}
   \clph = \frac43\pi^3 \left( \Mr{3}\sp\uone\sp\utwo\sp\uthree \right) + \equiv{u^5}. \label{eq:clph}
\end{equation}

\subsection{Resolving an object: size, asymmetry and curtosis}

\begin{figure*}[tp]
   \centerline{\includegraphics[width=0.7\hsize]{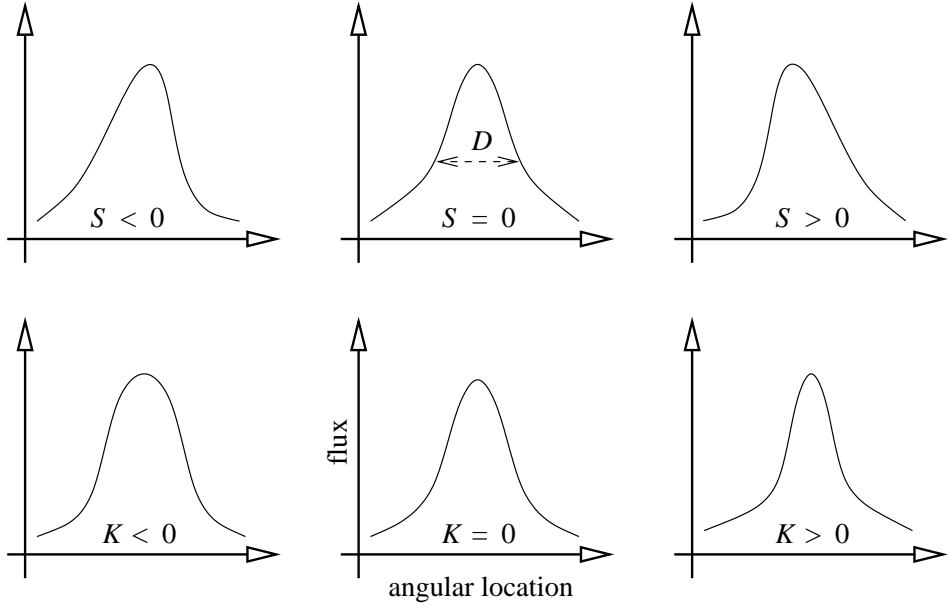}}
   \caption{%
      Link between the shape of a flux distribution and its first moments:
      mean diameter $D$, asymmetry coefficient $S$ (skewness), and 
      curtosis $K$.%
   }
   \label{fig:geom}
\end{figure*}

\begin{figure*}[tp]
   \begin{center}
   \parbox[c]{0.42\textwidth}{\includegraphics[width=0.42\textwidth]{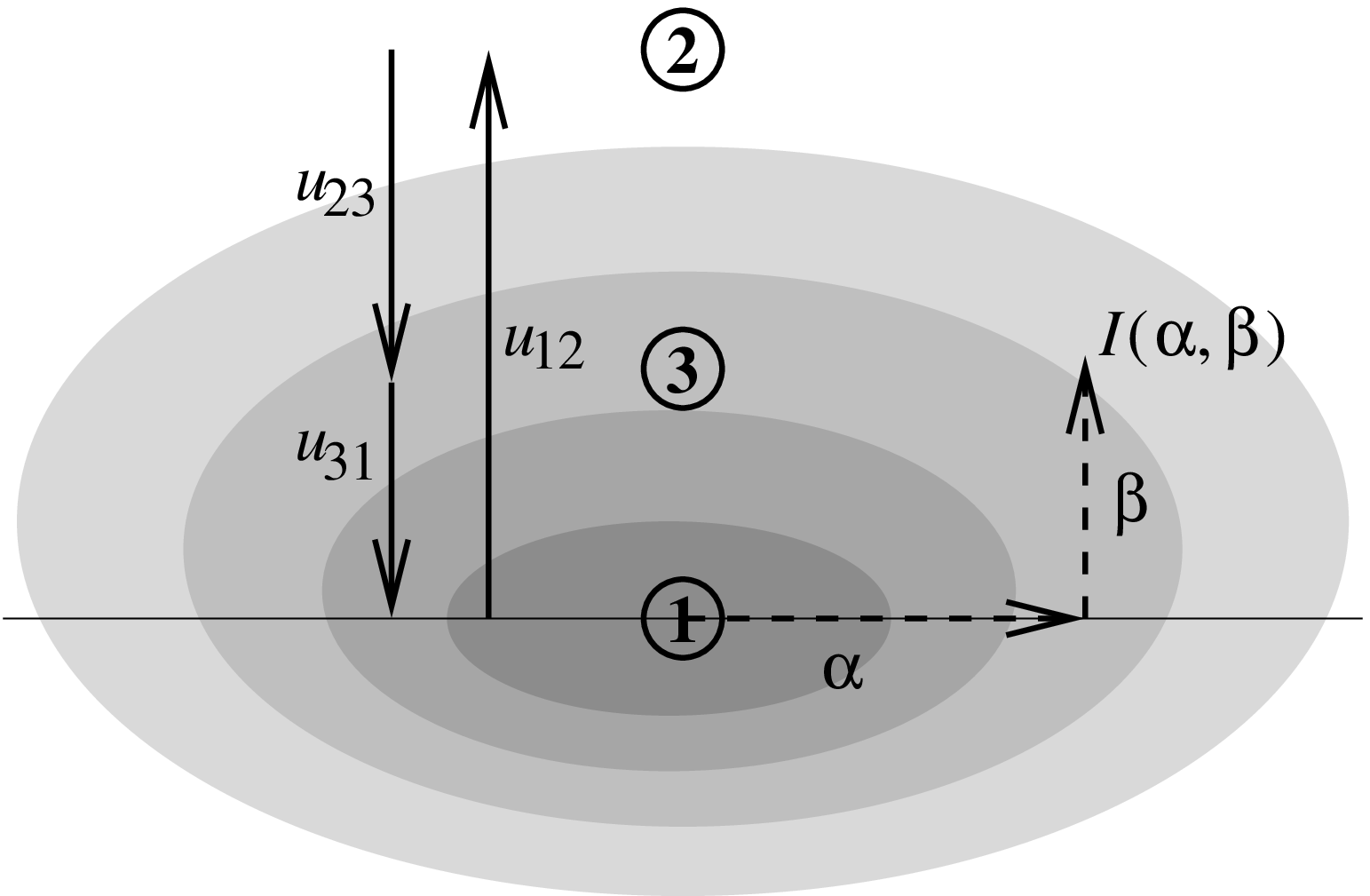}}
   \hglue 0.10\textwidth  
   \parbox[c]{0.52\textwidth}{%
      \begin{minipage}{0.52\textwidth}
      \large
      \begin{align*} 
         |V_{ij}|^2 &= 1 - 4\pi^2 u_{ij}^2 \underbrace{\iint I(\alpha, \beta) \beta^2 \idiff\alpha\idiff\beta}_{D^2}\\
         \clph      &= \frac43\pi^3 \underbrace{\vphantom{\iint} u_{12} u_{21} u_{31}}_{{\bar u}^3} \underbrace{\iint I(\alpha, \beta) \beta^3 \idiff\alpha\idiff\beta}_{S D^3}
      \end{align*}
      \end{minipage}%
   }
   \end{center}
   \caption{%
      \correct{%
      Link between the flux distribution $I(\alpha, \beta)$, the visibility 
      amplitude $|V|$,  and the closure phase $\clph$ for a marginally resolved
      object.  The  three aligned telescopes, numbered 1 to 3, and the 
      baselines are projected onto the sky.  Third-order terms and lower have
      been kept.}%
   }
   \label{fig:geom-sum}
\end{figure*}

\begin{figure*}[tp]
   \centerline{%
      \includegraphics[width=0.32\hsize]{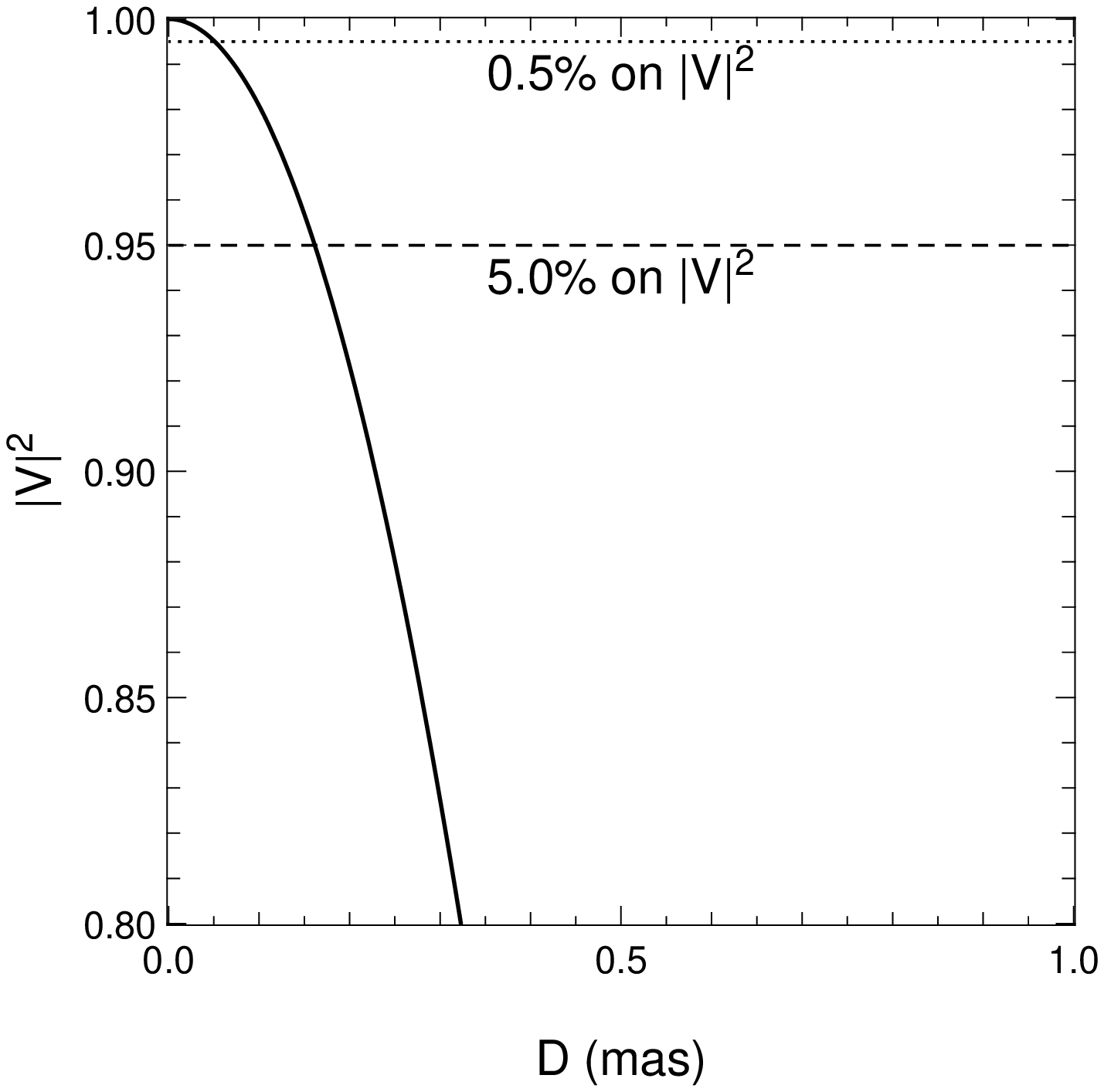}\hskip 0.01\hsize
      \includegraphics[width=0.32\hsize]{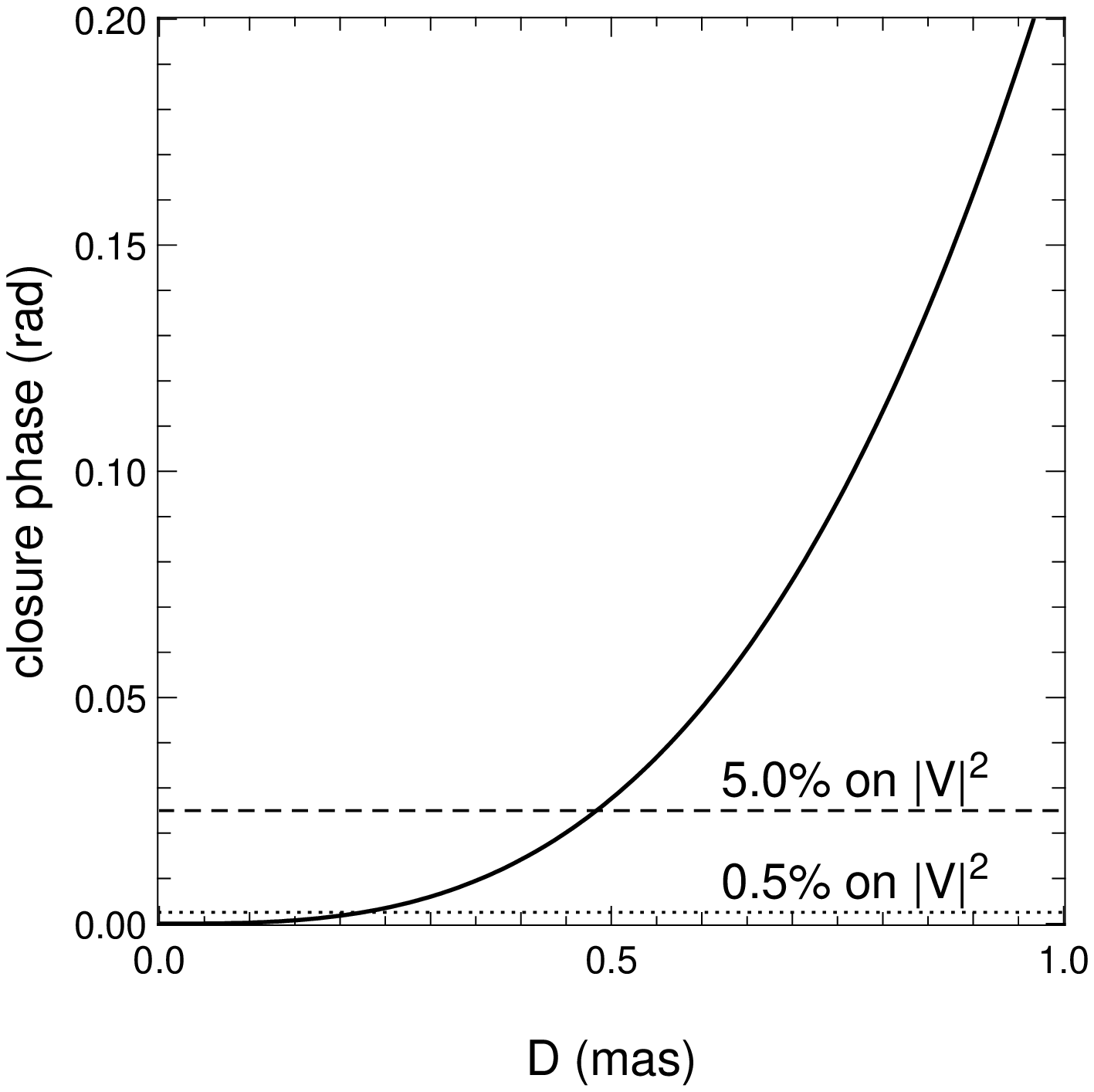}\hskip 0.01\hsize
      \includegraphics[width=0.32\hsize]{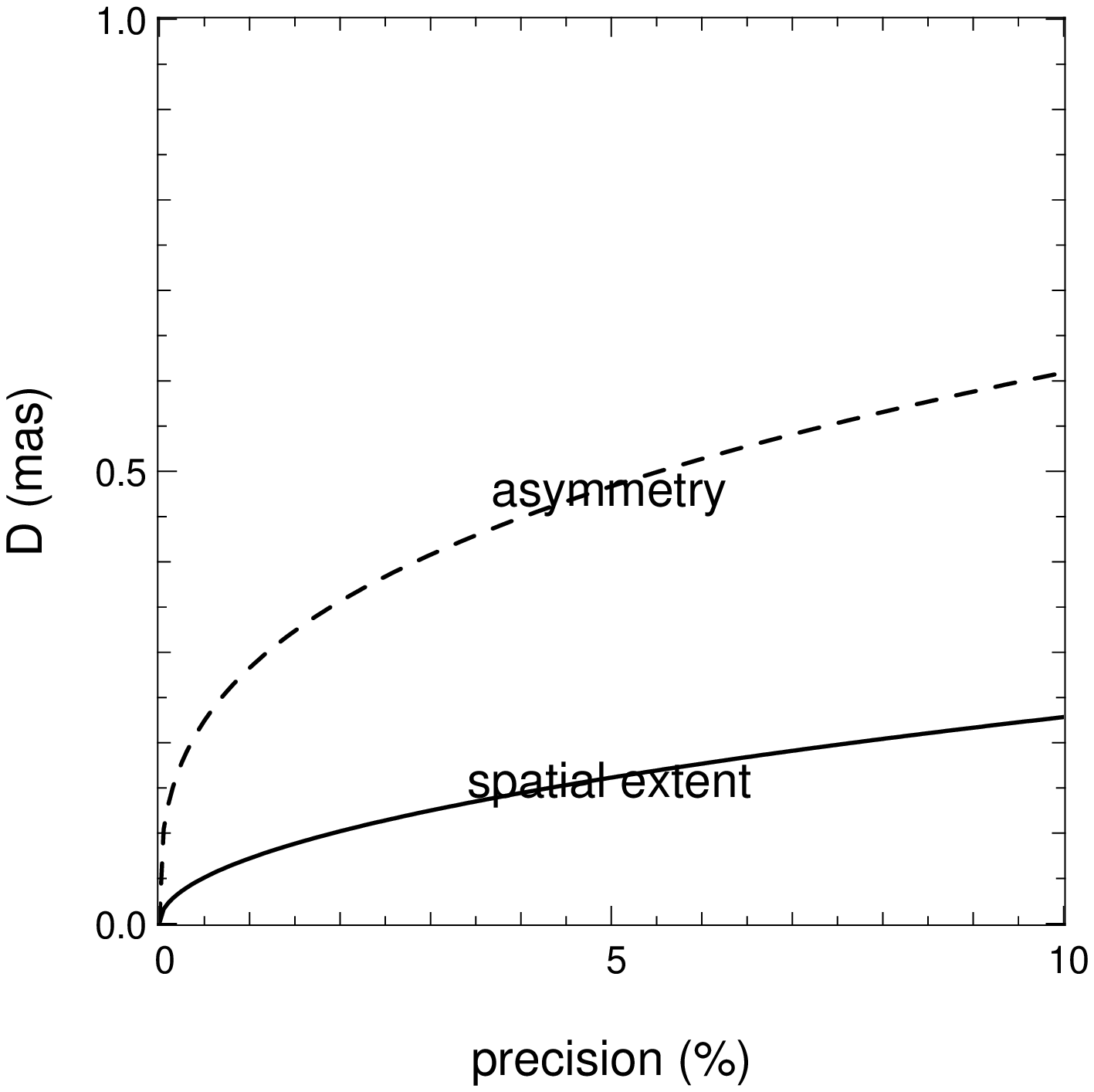}%
   } 
   \caption{%
      Variation of the square visibility amplitude and closure phase with the
      angular size of a marginally resolved object for a 100\,m baseline in K,
      and the detectability of these quantities.  Left panel: square visibility
      amplitude vs. angular size; middle panel: closure phase vs. angular size;
      right panel:  minimum object size needed to detect either the spatial
      extent or the asymmetry vs. the instrumental precision on $|V|^2$. On 
      the first left two view graphs the detection levels for 0.5\% and 5\% 
      \correct{accuracy on $|V|^2$ measurements} are displayed.  We assumed
      a fairly asymmetric object with $S = 0.5$, as well as
      $\Delta\clph = \Delta V$.
   }
   \label{fig:Vphi-prof}
\end{figure*}
We consider three aligned telescopes \correct{labelled from 1 to 3}, providing
the baselines $\uone$, $\utwo$, and $\uthree$ in a direction given by a normal
vector $\vi$, \correct{as represented in Fig.}~\ref{fig:geom-sum}.  The main
characteristics of an object we shall consider are its mean diameter, its
asymmetry and its curtosis defined along $\vi$;  they \textbf{respectively}
are
\begin{align}
   D &= \sqrt{\Mr{2}\sp\vi\sp\vi},\\
   S &= \frac{\Mr{3}\sp\vi\sp\vi\sp\vi}{\left(\Mr{2}\sp\vi\sp\vi\right)^{3/2}},\\
   K &= \frac{\Mr{4}\sp\vi\sp\vi\sp\vi\sp\vi}{\left(\Mr{2}\sp\vi\sp\vi\right)^2}-3.
\end{align}
They are of common use in statistics.  Figure~\ref{fig:geom} displays
their link with the shape of a one dimensional distribution:  the diameter $D$
is the square root of the variance, giving the mean spatial extent;  the
skewness $S$ increases with the asymmetry of the distribution and is zero for a
symmetrical one;  $K$ indicates whether the flux is concentrated in the peak of
the distribution or in its wings and is zero for a normal distribution.  With
these notations,
\begin{align}
   |V(u)|^2 &= 1 - 4\pi^2 (Du)^2 + \frac43\pi^4 \left[ (K+6)(D u)^4  \right], \label{eq:V-o4}\\
   \clph    &= \frac43\pi^3 S (D \bar{u})^3, \label{eq:clph-o4}
\end{align}
where the mean baseline $\bar{u}$ is given by the geometrical mean of
the three baselines:
\begin{equation}
   \bar{u} = \sqrt[3]{(\uone\sp\vi)(\utwo\sp\vi)(\uthree\sp\vi)}.
\end{equation}
A summary of these results, in a less formal way, is given in
Fig.~\ref{fig:geom-sum}.

The main implication of these results is that the visibility amplitude drop is
a second-order phenomenon ($D^2u^2$) while the closure phase is a third-order
one ($S D^3u^3$).  As a consequence, closure phase is much harder to detect than
the visibility amplitude in a marginally resolved object.
Figure~\ref{fig:Vphi-prof} displays the profile of the visibility amplitude and
closure phase as a function of the baseline for a marginally resolved object
with a high asymmetry $S = 0.5$, as well as the minima of detection for these
quantities as a function of the measurement accuracy.   It appears that the
asymmetry is detected for angular sizes \correct{3 to 6} times larger than the spatial
extent or ---which is equivalent--- for \correct{3 to 6} times larger baselines.

\subsection{Validity of the approximation}
\label{sec:validity}

\begin{figure*}[tp]
   \centerline{%
      \includegraphics[width=0.36\hsize]{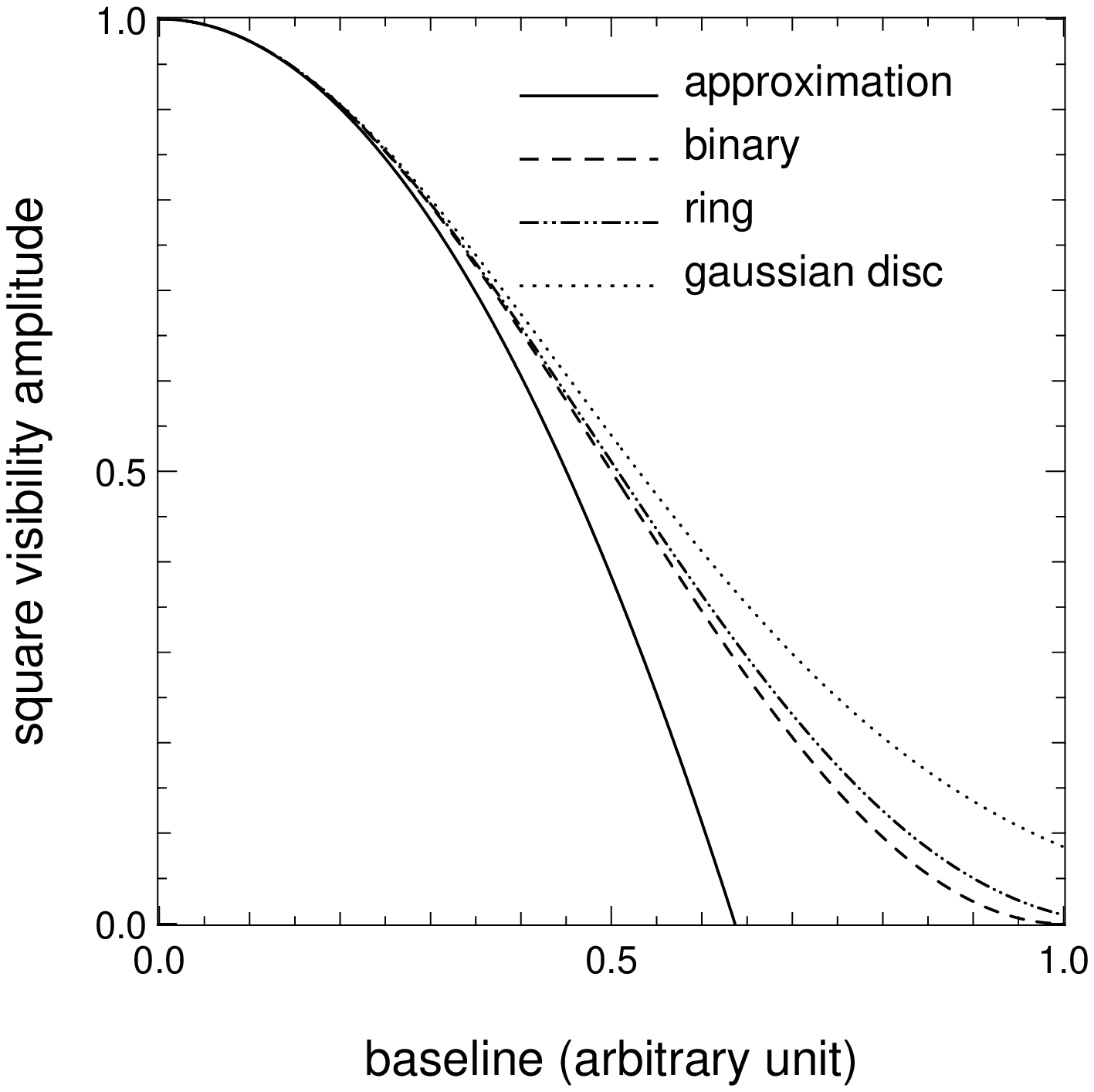}\hskip 0.1\hsize
      \includegraphics[width=0.36\hsize]{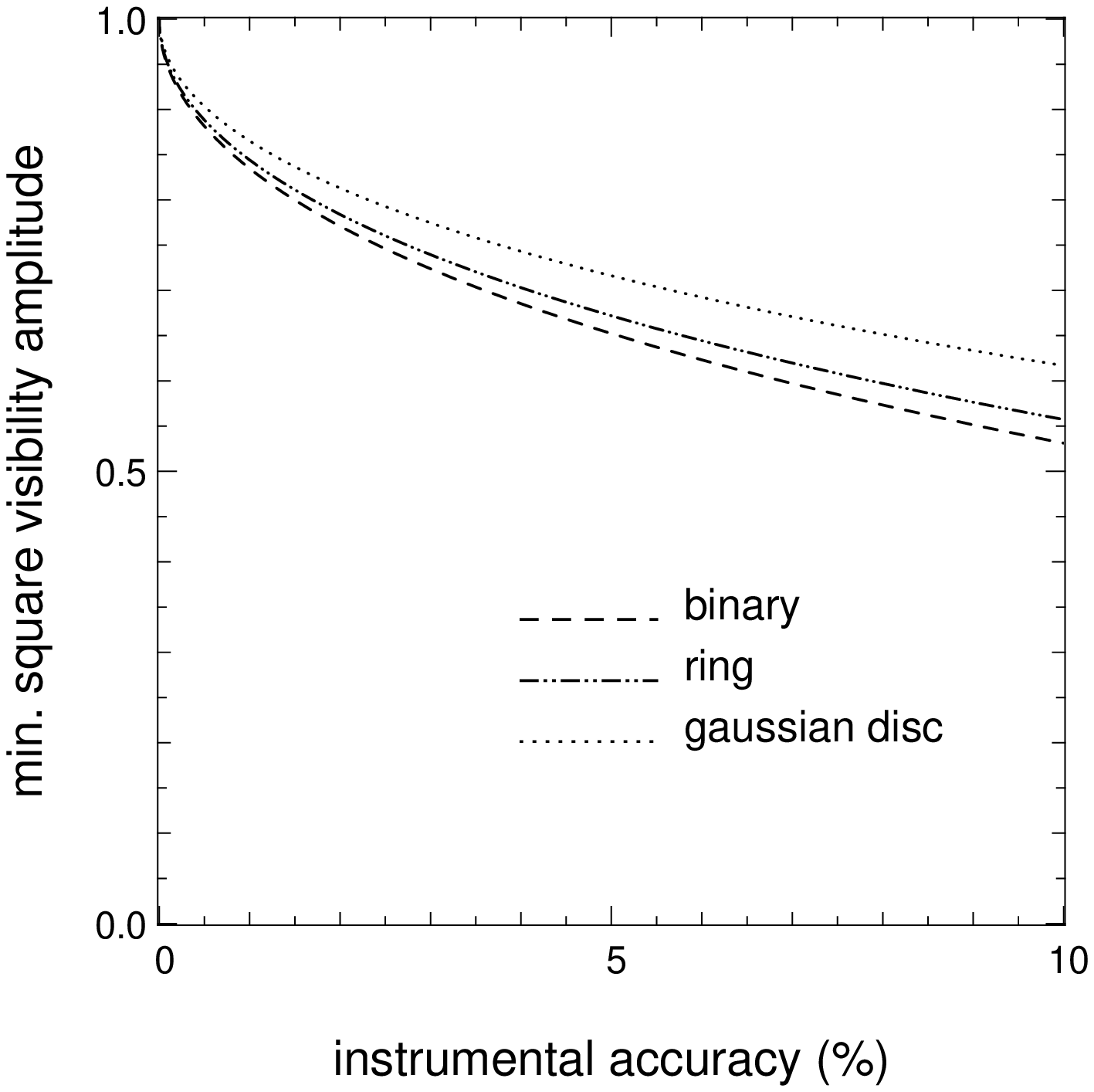}%
   }
   \caption{%
      Comparison between the exact visibility amplitude of an object
      and its second-order estimate.  \correct{Different geometries
      have been assumed to show that little dependence is found: a 
      symmetrical binary (dashes), a ring (dash-dot-dot), and a Gaussian disc 
      (dots).  The left panel displays the second-order estimate
      (solid line) and the exact visibilities of the different objects
      vs. the baseline.  The right panel displays the visibility amplitude
      for which an observational difference can be made between the
      estimate and the exact value vs. the instrumental precision;  it is 
      the limit of validity for this estimate.}
   }
   \label{fig:Vvalidity}
\end{figure*}

\begin{figure*}[tp]
   \centerline{%
      \includegraphics[width=0.36\hsize]{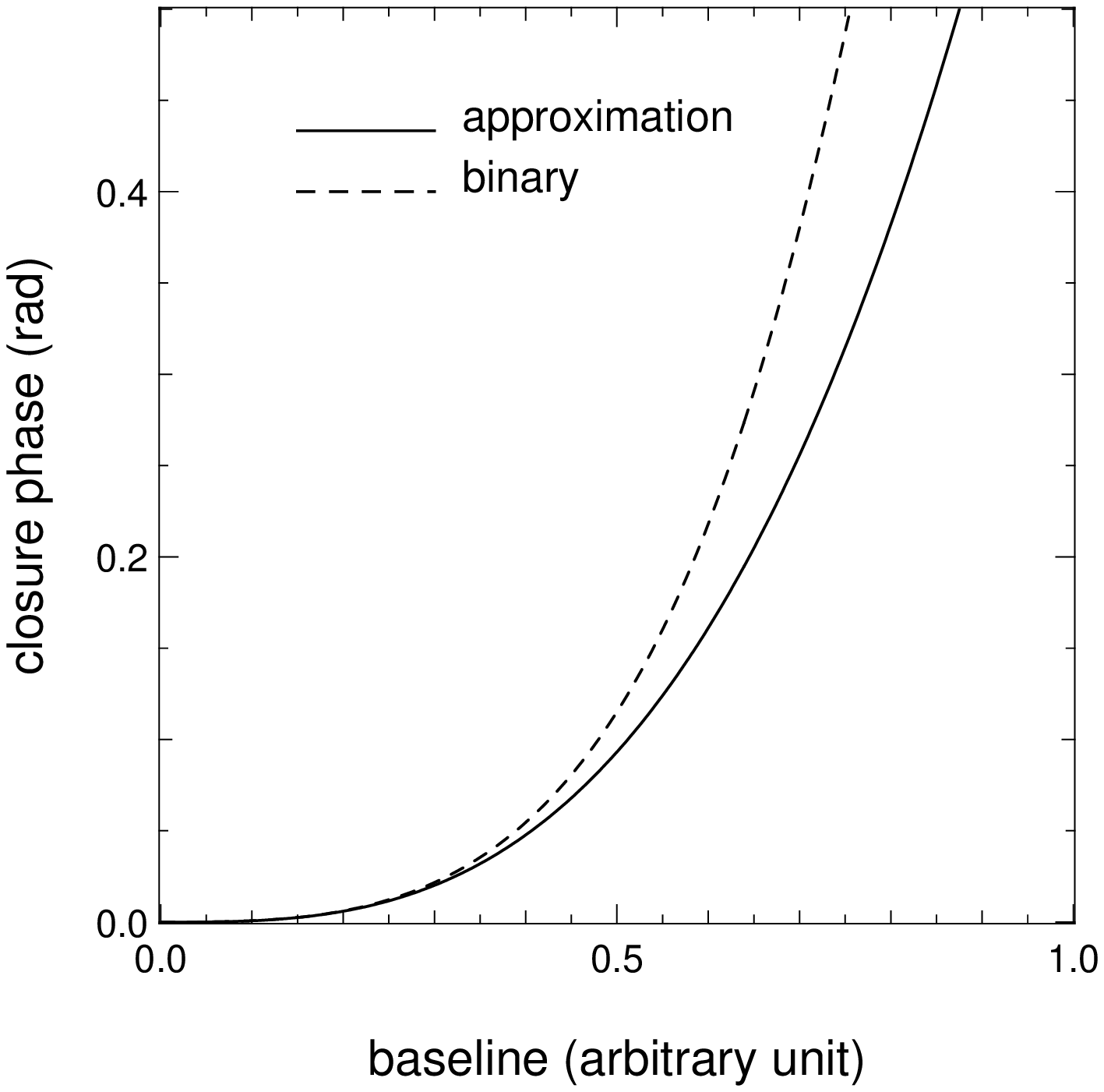}\hskip 0.1\hsize
      \includegraphics[width=0.36\hsize]{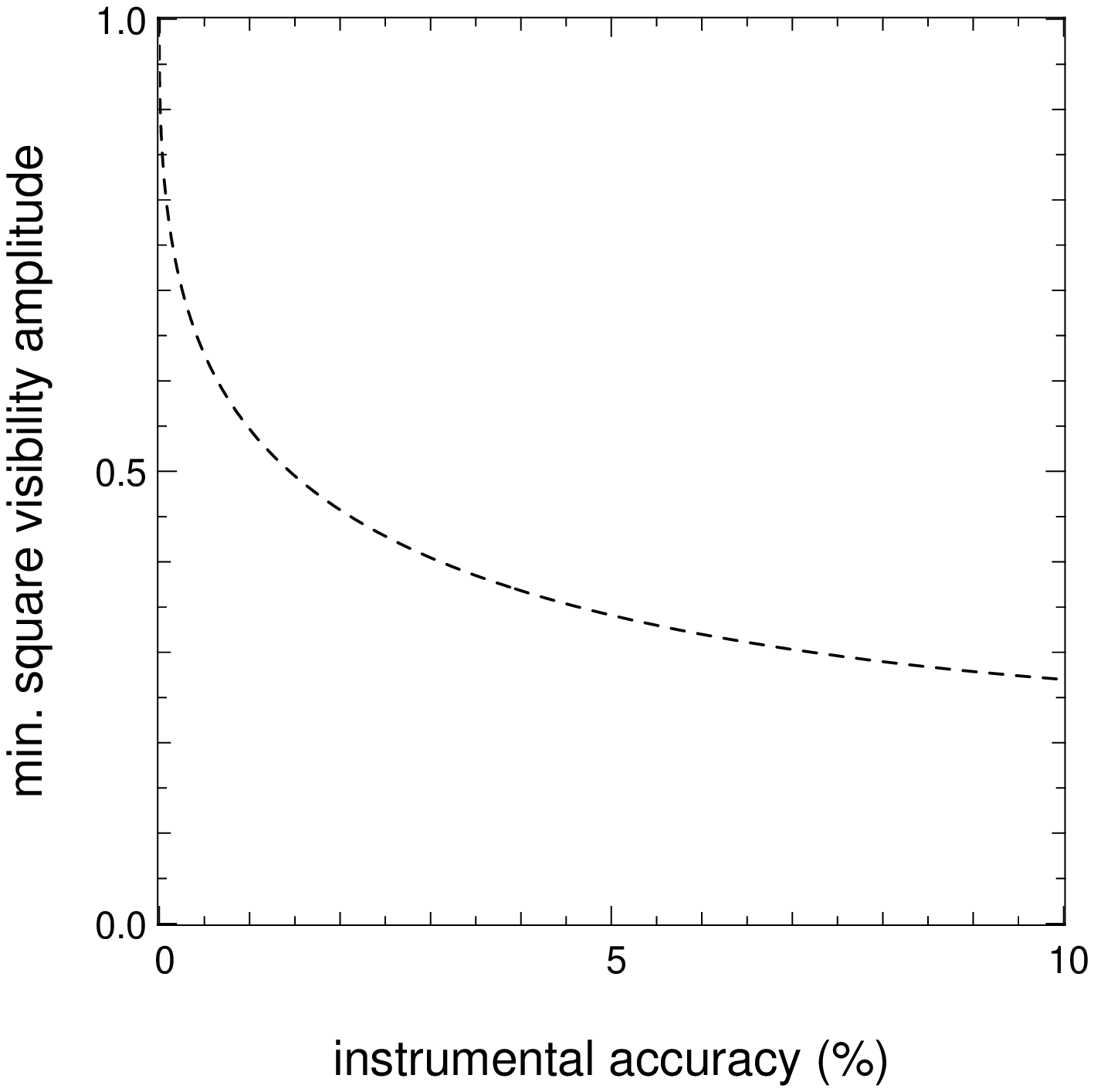}%
   }
   \caption{%
      Comparison between the exact closure phase of an asymmetrical binary
      ($\Delta\mbox{mag} = 1.2$) and its third-order estimate.  Left panel:
      closure phase vs. mean baseline.  Right panel: minimum visibility 
      for which no observational difference can be made between the estimate 
      and the exact phase vs. instrument accuracy.%
   }
   \label{fig:phivalidity}
\end{figure*}

The above development presents two limitations: on the one hand, it assumes
that all moments are defined and, on the other hand, the first few terms
of the series are no longer a good approximation when the object is
resolved enough. 

In the case of a power law distribution $I(\alpha) \propto \alpha^{-q}$, often
encountered when scattered light dominates,  the high-order moments are not
defined.  Therefore the above development is no longer valid.  For instance,
\citet{Lachaume02} have shown that a disc with scattering presents a quick drop of
the visibility amplitude near the origin $u = 0$, that definitely does not present
the smooth profile $|V|^2 = 1 - 4(\pi D u)^2$.   In Sect.~\ref{sec:separate},
we shall see how to treat scattering at a large scale, while using the above
formalism for other flux contributions.

Another important point is the range of validity.  The \correct{left panel} of
Fig.~\ref{fig:Vvalidity} compares the exact visibility of different types of
objects with the second-order estimate (first terms of Eq.~\ref{eq:V-o4}) as a
function of the baseline: \correct{as expected, the approximation is correct for
under resolved objects with $V \approx 1$ but gets poorer with larger
baselines.  The validity of the approximation depends on whether a difference can
be made between the estimate and the exact value, in other terms whether the
accuracy of the measurement is better than the precision of the estimate. The
right panel of the same figure displays the visibility at which the instrument
accuracy allows us to measure the deviation from the estimate; it appears not to
be much dependent on the geometry of the object.  With a typical 2\% accuracy
on $|V|^2$, the estimate is valid for $|V|^2 \gtrsim 0.8$.}

Figure~\ref{fig:phivalidity} is similar to Fig.~\ref{fig:Vvalidity} but for the
closure phase.  For a typical binary with $\Delta\mag = 1.2$, it displays the
exact closure phase and the third order estimate given in Eq.~(\ref{eq:clph-o4})
as a function of the baseline.  The right panel indicates the visibility at
which a differentiation between them can be made as a function of the instrumental
accuracy.  With the typical 2\% precision on $|V|^2$, the estimate of the phase
is valid for $|V|^2 \gtrsim 0.45$.

\correct{When the accuracy allows us to see the deviation from the third-order
estimates,  some following orders can be measured and the subsequent moments of
the distribution can be accessed.  This can be a means to retrieve
model-independent spatial information.}

\subsection{Model-fitting of observations}
\label{sec:modelfit}

\begin{figure*}
   \centerline{%
      \includegraphics[width=0.36\hsize]{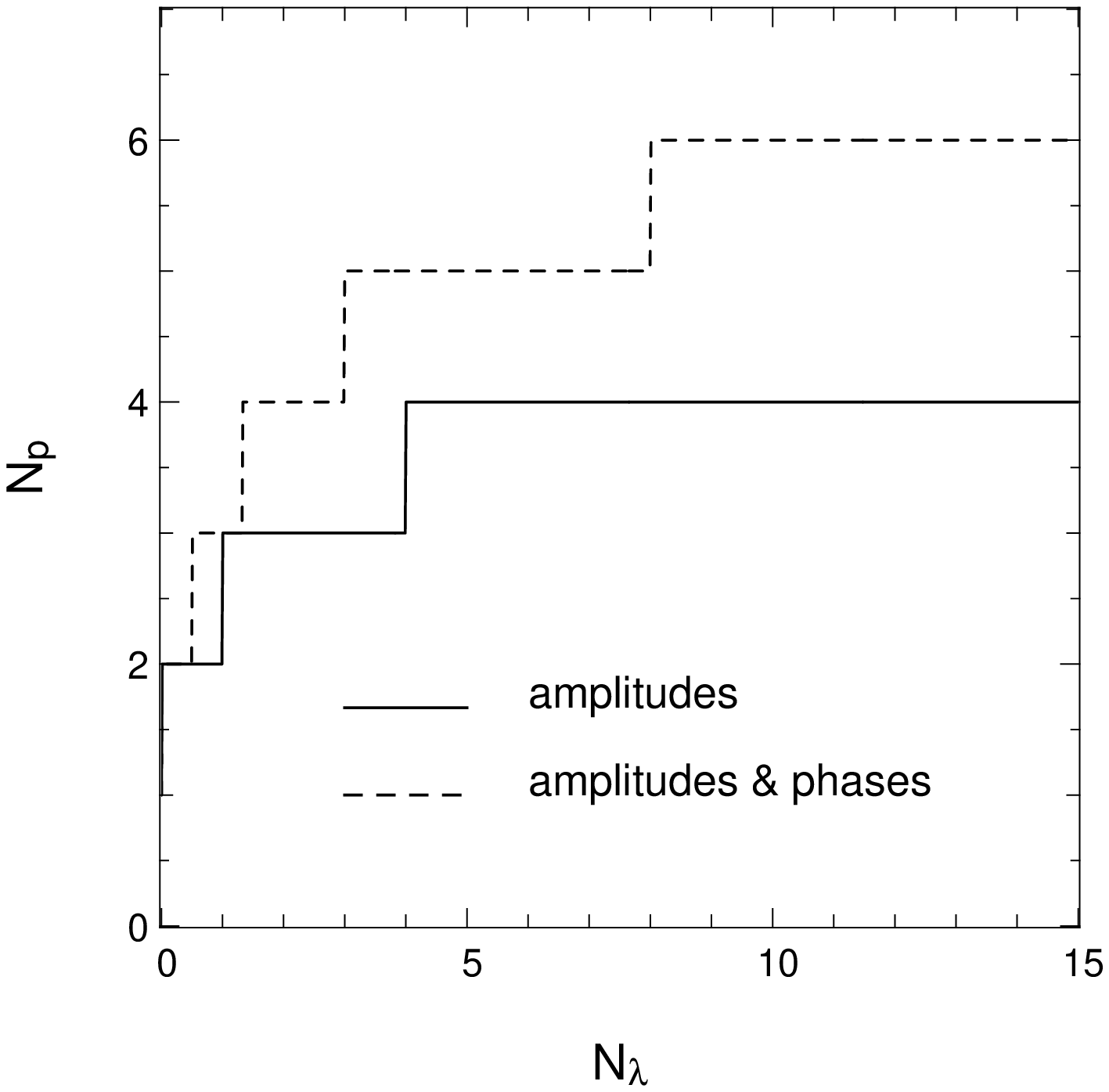}\hskip 0.1\hsize
      \includegraphics[width=0.36\hsize]{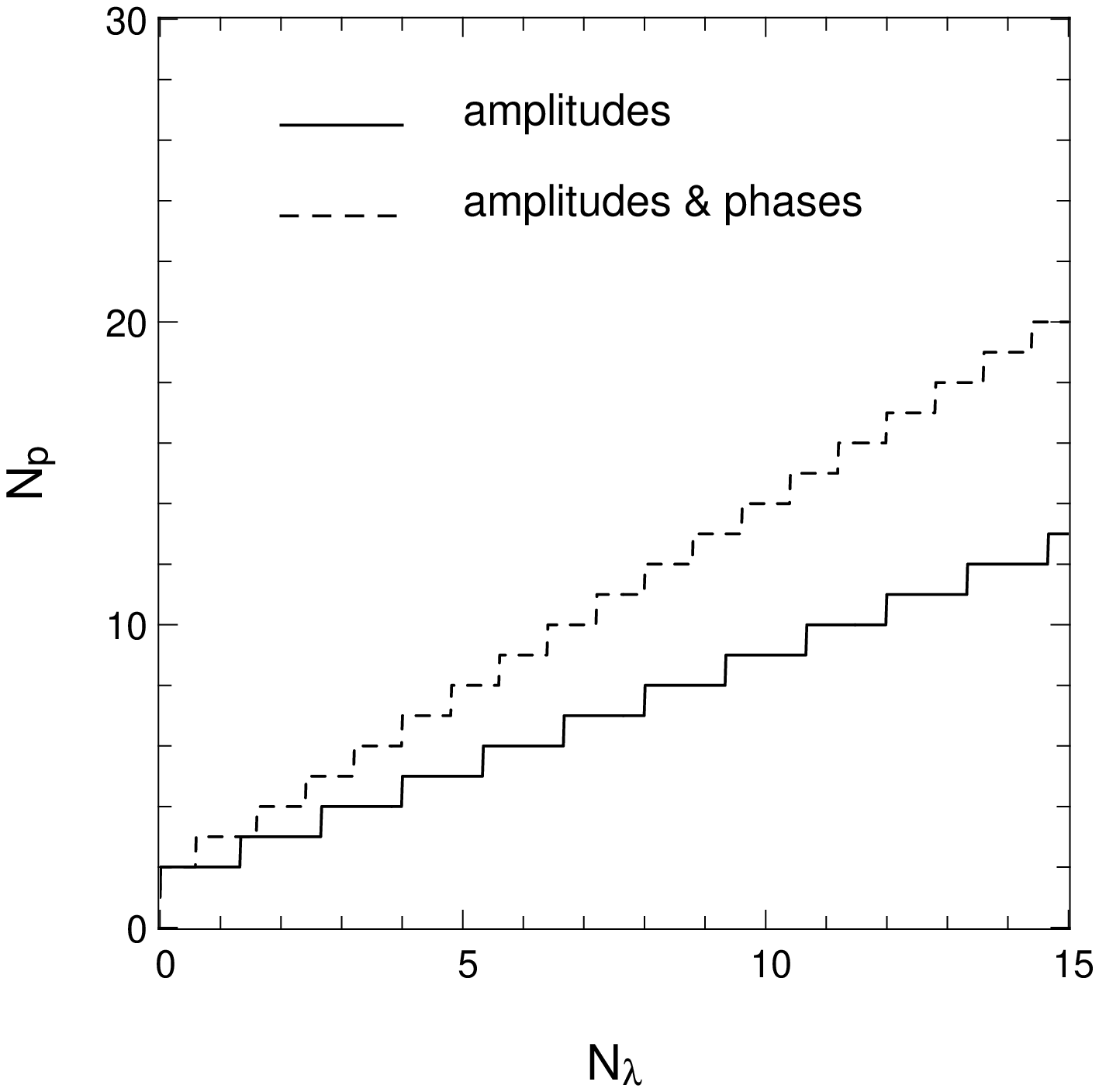}%
   }
   \caption{%
      Point-like source model fitting:  number of sources constrained by 
      interferometric observations of a marginally resolved object vs. number 
      of wavelengths used.  Left panel:  the fluxes at different wavelengths
      are not correlated.  Right panel:  the sources are assumed to emit
      a black-body spectrum.
   }
   \label{fig:constraints}
\end{figure*}

Theoretically speaking, the knowledge of visibility and closure phase at all
baselines smaller than $B$ would allow image reconstruction with an infinite
resolution (by using the analycity of the visibility); so, marginal resolution
should not be a problem.  This point presents a small but irretrievable flaw:
it assumes that there is no noise in the data.  When the finite precision of
measurements is taken into account, the accuracy on the reconstructed image or
model is impaired and also, the number of parameters constrained in a
marginally resolved object.  If the visibility amplitude alone is available,
only the quadratic form $V^2 = 1-4\pi^2 \Mr{2}\sp\vu\sp\vu$ can be accessed,
because the deviation from this law cannot be measured, as shown in
Sect.~\ref{sec:validity}.

We first consider an observation carried out at a single wavelength and use the
coordinates $(u, v)$ of $\vu$ in a Cartesian frame.   The visibility amplitude
and phase then are
\begin{align}
   V    &= 1 - 4\pi^2 (\M20 u^2 + 2\M21 uv + \M22 v^2),\\
   \phi &= 1 - \frac43\pi^3 (\M30 u^3 + 3\M31 u^2v + 3\M32 uv^2 + \M33 v^3),
\end{align}
where $\M nk$ is a component of the $n$-th moment $\Mr{n}$.  Its expression in
the Cartesian frame can be found in Appendix~\ref{ap:fm}.  If visibility
amplitude alone is considered, the system can be described with three
parameters (the $\M2k$); a model able to fit any second-order moment will
fit any data.  As a consequence, a marginally resolved object observed in
visibility amplitude, whatever the baseline coverage is, can be modelled by
\begin{itemize}
   \item a ternary system of unresolved stars,
   \item a Gaussian elliptic disc,
   \item a uniform stellar photosphere and an unresolved star.
\end{itemize}
If closure phase information is available, then four additional parameters
describe the object (the $\M3k$).  A marginally resolved object
observed in both visibility and closure phase can be modelled by
a system able to reproduce the second- and third-order moments, that is
\begin{itemize}
   \item a system of four unresolved stars,
   \item a Gaussian elliptic disc and an unresolved star,
   \item a uniform stellar photosphere and two unresolved stars.
\end{itemize}
It appears that interferometry does not allow us to disentangle quite
different scenarios when the object is marginally resolved.  \correct{This fact
is well-known by observers:  under resolved observations cannot distinguish
between
a uniform and a limb-darkened stellar disc; neither could the first
observations of FU~Ori by} \citet{Malbet98} \correct{exclude either the disc or
the binary scenario.}  As a remedy, one can use the technique in combination
with other types of observations, as another model constraint.  

Nevertheless, a multi-wavelength interferometric approach can bring more
constraints.  We consider a $\Np$ point-like source model to be fitted to
visibility amplitudes at $\Nlambda$ wavelengths.  The fluxes and locations of
the $\Np-1$ first sources constrain the last one, because the flux distribution
is normalised and centered; therefore there are $(\Np-1)$ locations and
$\Nlambda (\Np-1)$ fluxes, that is $(\Nlambda+2)(\Np-1)$ free parameters.
Observations provide $3\Nlambda$ moments of the flux distribution.  The
characteristic number of point-like sources constrained by the measurements is
given by the equality between the number of free parameters and that of
moments, so that
\begin{equation}
   \Np = \left\lceil \frac{3\Nlambda}{\Nlambda+2}+1 \right\rceil.
\end{equation}
The left panel of Fig.~\ref{fig:constraints} displays the number of sources
constrained as a function of the number of wavelengths accessed, either
observing visibility amplitudes only or both amplitudes and closure phases.
It appears that amplitudes and phases can constrain up to 6 point-like sources
with $\Nlambda \gtrsim 10$.  If we now consider that the sources
emit a black-body spectrum ---or whatever spectrum determined
by a temperature and a bolometric flux--- there are only $2(\Np-1)$ fluxes. 
The number of point-like sources constrained then becomes
\begin{equation}
   \Np = \left\lceil \frac{3\Nlambda}{4}+1 \right\rceil.
\end{equation}
The right panel of Fig.~\ref{fig:constraints} shows that the assumption of a
black-body spectrum allows us to constrain as many point-like sources as wanted,
provided that the number of wavelengths is large enough.  For instance,
$\Nlambda = 10$ allows us to constrain 9 sources with the visibility amplitudes
and 14 if phases are also available.  

In conclusion, a multi-wavelength approach allows us to fit more 
parameters and therefore to distinguish between different scenarios (disc, multiple 
system, etc.)


\section{Application to circumstellar discs}
\label{sec:disc}

\begin{figure}[t]
   \centerline{\includegraphics[width=0.75\hsize]{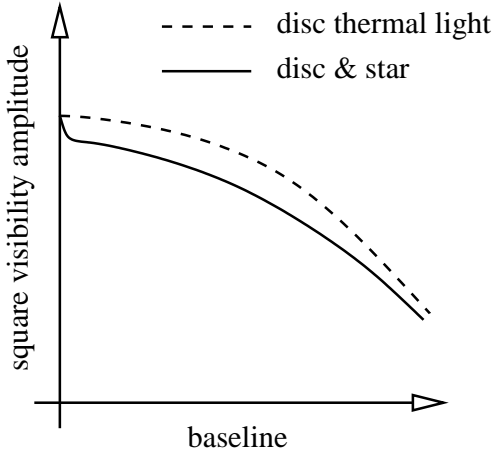}}
   \caption{%
      Visibility curve of an accretion disc. Solid line: all contributions;
      dashed line: thermal emission of the disc only.%
   }
   \label{fig:diskvisib}
\end{figure}

Circumstellar discs are a good target for interferometers because they scale
from a few tens of AU (where they mostly emit thermal light in the infrared) to
a few hundreds of AU (where they present scattered light in the infrared) at a
typical distance of 150\,pc or more.  Two issues of interest are:  observing
their thermal light, because it comes from the first AUs from the star where
planets are supposed to form, and deriving their radial temperature law,
because it appears as a good diagnosis of the phenomena involved (irradiation,
flaring, viscous dissipation, etc.).  In Sect.~\ref{sec:separate} we show how
to take into account both the thermal and scattered light, which happen to
present different interferometric signatures.  In Sect.~\ref{sec:powerlaw}, we
establish a connection between the temperature law and the wavelength
dependence of the visibility.

\subsection{Stellar, thermal and scattered light}
\label{sec:separate}

Describing an accretion disc as a marginally resolved object is inaccurate
because thermal light occurs at a large scale and accounts for up to 10\% of
the total flux.  In order to keep the above formalism, we split the image into
three components: stellar contribution, thermal emission of the disc, and
scattered light.  To each contribution, one can associate a corresponding
visibility:
\begin{align}
   \Vstar    &= 1 - 2\pi^2 \Dstar^2 u^2,\\
   \Vtherm   &= (1 - 2\pi^2 \Dtherm^2 u^2) \exp\left( i \frac43\Stherm\Dtherm^3 u^3 \right),\\
   \Vscat    &=
      \begin{cases}
         1 &\quad\mbox{if $u = 0$}\\
         0 &\quad\mbox{otherwise},
      \end{cases}
\end{align}
where $\Dstar$ is the mean diameter of the star, $\Dtherm$ that of the
thermal emission of the disc, and $\Stherm$ the skewness of distribution
of the thermal emission.

We assumed that both the star and the thermal emission of the disc are marginally
resolved, that the scattering emission is fully resolved as soon as the
baseline is non-zero, and that the star is too small and symmetric to present a
phase.  Within the approximation that all components have the same photocentre, 
we derive the total visibility
\begin{equation}
   \Vtot = \frac{\Fstar\Vstar + \Ftherm\Vtherm + \Fscat\Vscat}{\Ftot},
\end{equation}
where $\Ftot$ is the total flux, $\Fstar$ that of the star, $\Ftherm$ that
of the thermal emission, and $\Fscat$ that of scattered light.  
Fig.~\ref{fig:diskvisib} presents a schematic view of the visibility
curve for an accretion disc and compares it to that of the thermal
light alone.  It appears that one point of visibility is not enough to
derive the diameter of the disc, as authors usually do, when either the
star or scattered flux are present.

\subsection{Temperature profile}
\label{sec:powerlaw}

From visibilities at different wavelengths, \citet{Malbet02p} showed that
the temperature profile of an accretion disc can be derived.   In the
context of a massive disc, the flux is dominated by thermal light so
that
\begin{equation}
   1-|V(\lambda)|^2 = \left( 2\pi D(\lambda) B / \lambda \right)^2.
\end{equation}
To solve this problem, we need a link between the disc extent $D(\lambda)$ and
the temperature profile.  The disc extent is given by the second-order moment
$\Mr{2}$ determined from the radial flux distribution of the disc
\begin{equation}
   F(r) = B_\lambda\left(T(r)\right)\label{eq:F(r)}.
\end{equation}
We chose a self-similar solution for the sake of simplicity
\begin{equation}
   T(r) \propto r^{-q}.
\end{equation}
For a disc presenting an inclination $i$, the components of the moments
then write (see Appendix~\ref{sec:momss} for a demonstration):
\begin{subequations}
\begin{align}
   \M20 &=       \M22 \cos^2 i,\\
   \M21 &=       0,\\
   \M22 &\propto \lambda^{2/q},
\end{align}
\end{subequations}
so that 
\begin{align}
   D(\lambda) &\propto \lambda^{2/q},\\
   1-|V|^2    &\propto \lambda^{2/q-2}.
\end{align}
Using two close enough wavelengths $\lambda_1$ and $\lambda_2$, we deduce an 
estimate of $q$:
\begin{equation}
   q = \frac{1}{\displaystyle 1+ \frac12 \frac{\log [(1-|V_1|^2)/(1-|V_2|^2)]}{\log [\lambda_2/\lambda_1]}}.
   \label{eq:q}
\end{equation}
This result appears as a particular case of the multi-wavelength approach (see
Sect.~\ref{sec:modelfit}), that allows us to fit more parameters than one
visibility at a single wavelength does. \citet{Malbet02p} obtained a similar
result by a more empirical argument: they state that the apparent diameter of a
disc is proportional to the radius at which $B_\lambda (T(r))$ is maximal.  It
actually conceals the self-similarity.

\begin{figure*}[t]
   \centerline{%
      \includegraphics[width=0.36\hsize]{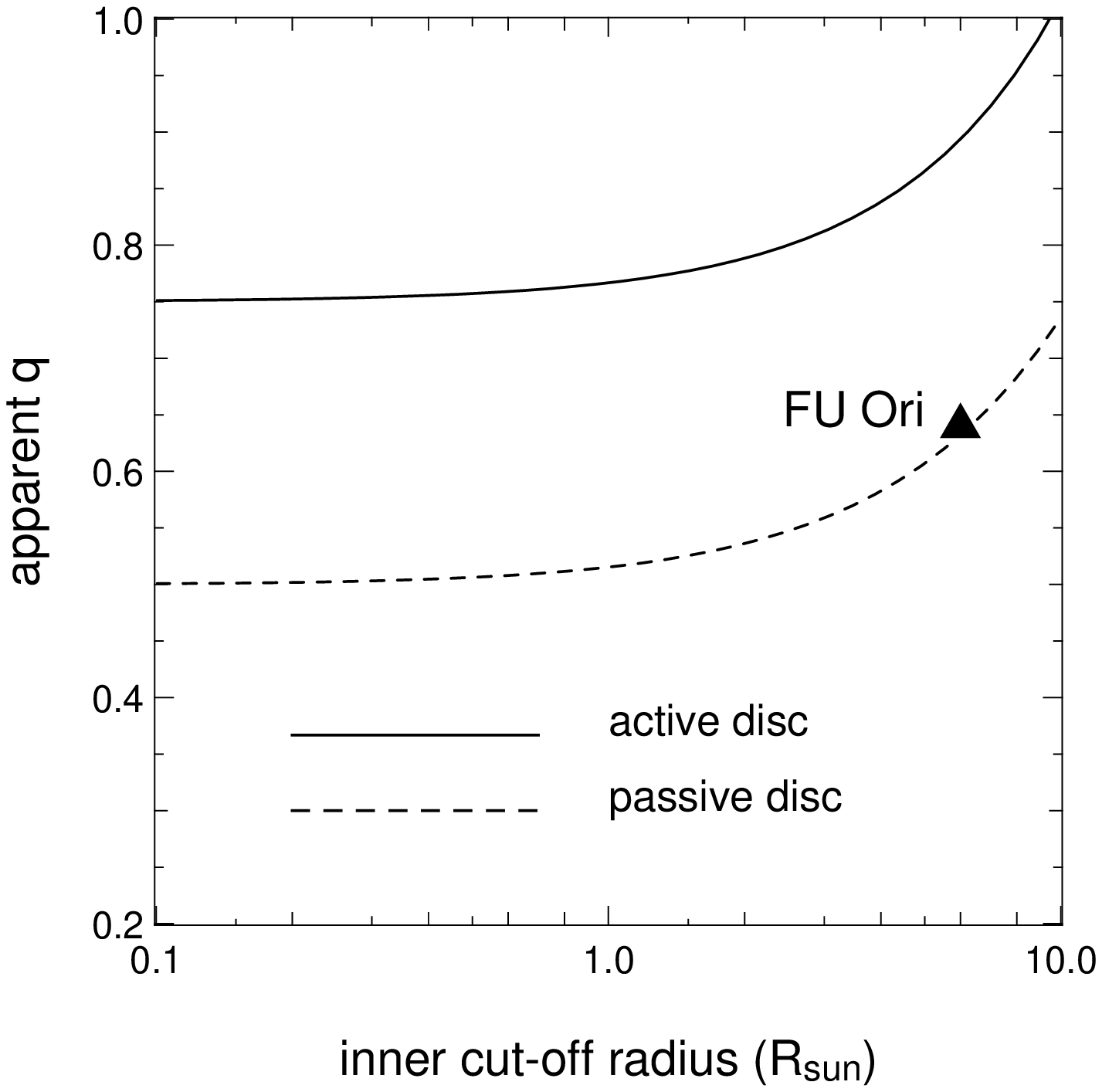}\hskip 0.10\hsize%
      \includegraphics[width=0.36\hsize]{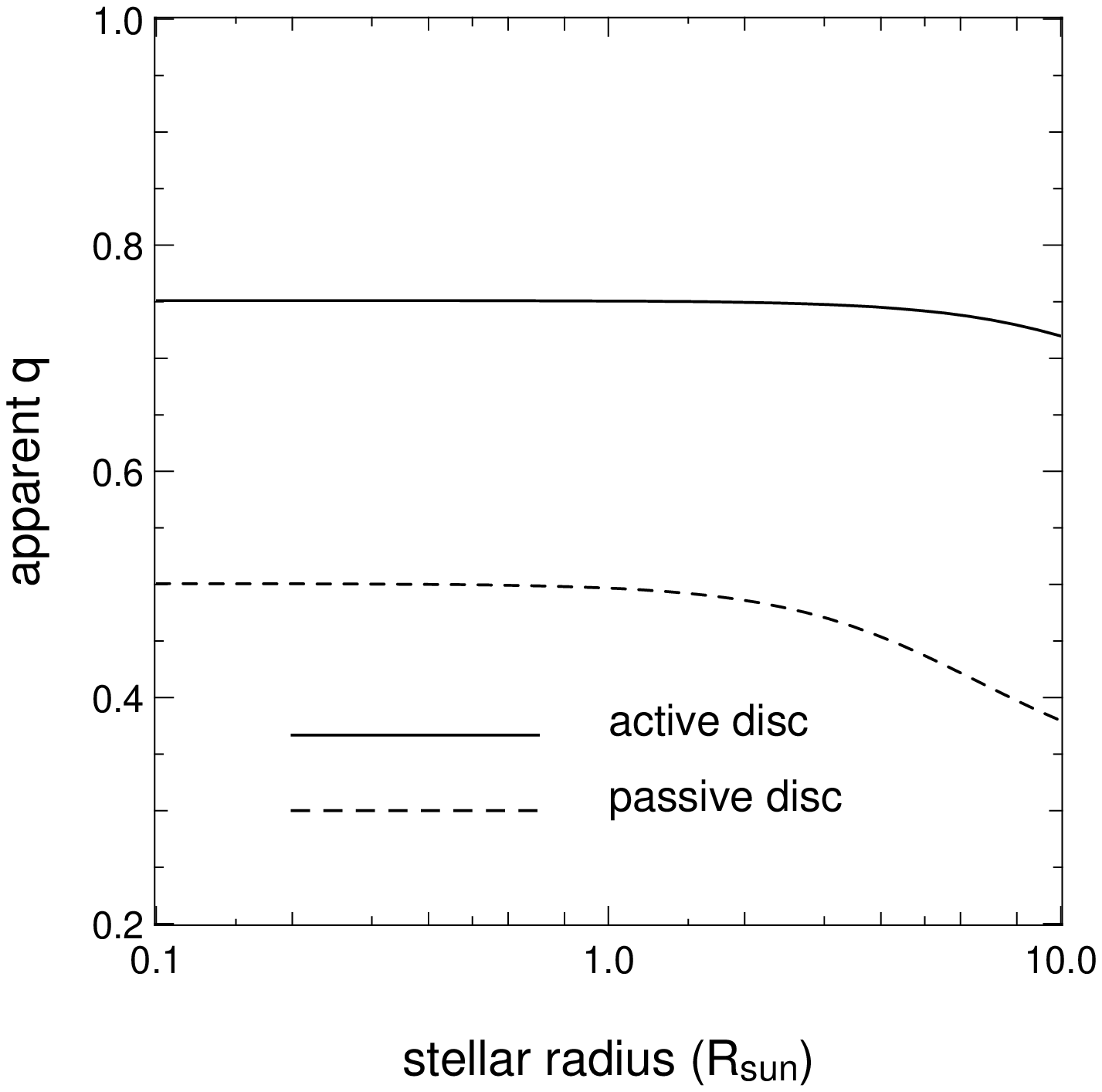}%
   }
   \caption{%
        Apparent temperature law, given by the exponent $\qapp$ as a function
        of the disc inner truncation (left) and the stellar radius (right).  
        Solid line: active disc with $q = 0.75$ and an accretion rate of
        $3\times10^{-5}\,\Mdot/\yr$; dashed line:  flared passive disc with 
        $q = 0.5$ and the same effective temperature at 1\,AU.  The star
        is a black-body at temperature 5000\,K.  The two visibilities are 
        taken in H and K.
   }
   \label{fig:qapp}
\end{figure*}

Note that the result no longer holds when the disc presents an inner
hole.  Figure~\ref{fig:qapp} displays the value $\qapp$ deduced from
Eq.~(\ref{eq:q}) with the H and K bands for a typical FU~Ori disc.  The
parameters of the disc model are given in Appendix~\ref{sec:pladm}.
When the inner gap becomes larger than a few stellar radii, the error on $q$,
$|\qapp-q|$ can be larger than $0.1$.  \citet{Malbet02p} find $\qapp \approx 
0.64$ for FU~Ori; with a typical value $R_* = 6\,\Rsun$ \citep{Lachaume02}, 
we can estimate $q \approx 0.5$ from the curves.   

The stellar radius has also an influence because of the unresolved stellar
flux, yet, it remains small for FU~Ori discs (see Fig.~\ref{fig:qapp}).  In the
case of a T~Tauri star, the error could be much larger, because the
contribution of the star to the total flux becomes important.


\section{Conclusion}

We have developed a formalism that connects the visibility amplitude and phase
of a marginally resolved object with its geometry, namely the moments of the
flux distribution.  It can prove particularly useful when constraining models
that present analytical moments \correct{and allows us to retrieve
model-independent spatial information in all cases}.  It also establishes that
the closure phase of a marginally resolved source is a third-order term and the
visibility a second-order one; therefore, the phase is much harder to detect
than the drop in visibility amplitude.

From the formalism, we were also able to estimate the number of parameters
relevantly fitted to interferometric measurements.  Unless observations are
carried out at several wavelengths and the model assumes a black-body-like
emission, only a few parameters can be fitted to marginally resolved objects,
whatever the number of visibility points taken:  three point-like source
with visibility amplitudes only, and a fourth one if closure phase is also
measured. \correct{This limitation is removed when the object is more resolved, that is,
if the baselines are longer or if the instrumental accuracy is increased,
which allows us to measure the deviation of the visibility and phase from
their low-order estimates.  This work can therefore be seen as a plea
for larger baselines than the CHARA array provides, or high accuracy
with IOTA/IONIC or the forthcoming VLTI and Keck.}

We then applied this theoretical work to circumstellar discs, by separating the
star, the thermal emission of the disc, and the scattered light, the two first
ones being well described by their moments.  It also allows us to derive, with
some hypotheses, information on the radial temperature law in these objects,
\correct{even if they are underresolved}, but requires that measurements
should be taken at two \correct{or more} wavelengths. \correct{This
can be applied to other field.  For instance, limb-darkened stellar 
photospheres can be probed even with underresolved targets:  the
equivalent diameter is dependent on the wavelength and one could, with an
appropriate model, measure this darkening.}

\correct{With high precision measurements and/or multiple wavelengths one can
access a large number of moments of the flux distribution, which theoretically
allows image reconstruction.  This is clearly a path that one should
investigate in the near future.  As a particular case, we believe it is possible
to retrieve the radial temperature profile of supposedly symmetrical objects,
as as been initiated with FU~Ori.  The method could also prove useful
to constrain the location of stellar spots with high accuracy measurements:
the link between the location of these spots and the first order moments
of the flux is much clearer than the information given by image reconstruction 
techniques.}


\begin{acknowledgements}
I thank Fabien Malbet, Jean-Philippe Berger, and Jean-Baptiste Lebouquin for 
helpful discussions;  without their interest, this work would have 
stayed hidden under a heavy stack on my desk.   Computations and graphics have
mostly been carried out with the free software Yorick by D. Munro.
\correct{Useful comments from Theo ten Brummelaar led to an improved
presentation of these results.}
\end{acknowledgements}


\appendix

\section{Moments in Cartesian coordinates}
\label{ap:fm}

We use a Cartesian frame in which $\valpha$ has coordinates $(\alpha, \beta)$ 
throughout this appendix.

The first moment is a vector 
\begin{equation}
   \Mom 1 = (M_1^0, M_1^1)
\end{equation}
giving the location of the photocentre in respect to the origin of the
frame.  
\begin{subequations}
\begin{align}
   M_1^0 &= \iint I(\alpha, \beta) \alpha \idiff\alpha\idiff\beta,\\
   M_1^1 &= \iint I(\alpha, \beta) \beta  \idiff\alpha\idiff\beta.
\end{align}
\end{subequations}

The second-order moment is a matrix
\begin{equation}
   \Mom 2 = \left(
   \begin{matrix}
      M_2^0 & M_2^1\\
      M_2^1 & M_2^2
   \end{matrix} \right)
\end{equation}
yielding information on the spatial extent of the object.  Its components
are
\begin{subequations}
\begin{align}
   M_2^0 &= \iint I(\alpha, \beta) \alpha^2    \idiff\alpha\idiff\beta,\\
   M_2^1 &= \iint I(\alpha, \beta) \alpha\beta \idiff\alpha\idiff\beta,\\
   M_2^2 &= \iint I(\alpha, \beta) \beta^2     \idiff\alpha\idiff\beta.
\end{align}
\end{subequations}

The $n$-th order moment is a tensor of order $n$ containing $2^n$ elements. 
Since it is symmetrical it is given by only $n+1$ terms, that we note $M_n^k$
for $0 \le k \le n$:
\begin{equation}
   M_n^k = \iint I(\alpha, \beta) \alpha^{n-k} \beta^k \idiff\alpha\idiff\beta.
\end{equation}

\section{$V$ and $\phi$ series development}
\label{sec:hod}

We define 
\begin{align}
   J_1 &= \Mom{n} \sp\vu,\\
   J_n &= \Mr{n} \sp \underbrace{\vu \cdots \vu}_{\mbox{\scriptsize $n$ times}}, \quad\mbox{for $n \ge 2$}.
\end{align}
The visibility amplitude and phase are then expressed as:
\begin{gather} 
   \begin{split}
      V &=     1-4\pi^2 \,J_2 +\frac{4\pi^4}{3}  \left(J_4 + 3 J_2^2\right)\\
        &\quad  -\frac{8\pi^6}{45} \left(J_6 - 10 J_3^2 + 15 J_4 J_2\right)\\
        &\quad  +\frac{4\pi^8}{315}\left(J_8 + 35 J_4^2 - 56 J_3 J_5 + 28 J_6 J_2\right)\\
        &\quad  +\equiv{u^{10}},
   \end{split}\\
   \begin{split}
      \varphi &=  2\pi\, J_1 -\frac{4\pi^3}{3}   \,J_3 +\frac{4\pi^5}{15}  \left(J_5-10 J_3 J_2\right)\\
                &\quad -\frac{8\pi^7}{315} \left(J_7 + 219 J_3 J_2^2 - 21 J_2 J_5 - 35 J_3 J_5 \right)\\
                &\quad +\frac{4\pi^9}{2835}\left(J_9 + 2520 J_3 J_2 J_4 + 560 J_3^3  - 36 J_7 J_2 \right.\\
                &\qquad \left. - 84 J_3 J_6 - 126 J_5 J_4 + 756 J_5 J_2^2 - 7560 J_3 J_2^3\right)\\
                &\quad +\equiv{u^{11}}.
   \end{split}
\end{gather}

\section{Moments of a self-similar disc}
\label{sec:momss}

We consider a face-on disc with a radial temperature law $T(r) \propto r^{-q}$,
where $r$ is the angular distance from the centre.  The flux distribution
then reads
\begin{equation}
   F(r) = \frac{C_1}{\lambda^5 \left( \exp \frac{C_2}{\lambda r^{-q}} - 1\right) }, 
\end{equation}
where $C_1$ and $C_2$ are constants.  The reduced moments are
\begin{align}
  \M20 &= \frac 1F \iint F(x, y) x^2 \idiff{x}\idiff{y}, \label{eq:ap:m20}\\
  \M21 &= \frac 1F \iint F(x, y) xy  \idiff{x}\idiff{y}, \label{eq:ap:m21}\\
  \M22 &= \frac 1F \iint F(x, y) y^2 \idiff{x}\idiff{y}, \label{eq:ap:m22}
\end{align}
where $F$ is the total flux given by
\begin{equation}
   F = \iint F(r) 2\pi r \idiff{r}.
\end{equation}
For the sake of simplicity, we introduce the integral 
\begin{equation}
   I_s = \int F(r) 2\pi r^s \idiff{r} \label{eq:ap:Iq}.
\end{equation}
By switching to polar coordinates in Eq.~(\ref{eq:ap:m20}) +
Eq.~(\ref{eq:ap:m22}) and with the help of symmetry, we derive 
\begin{align}
  \M20 &= I_3/(2I_1),\\
  \M21 &= 0,\\
  \M22 &= I_3/(2I_1).
\end{align}
We perform the change of variables $u = C_2/(\lambda r^{-q})$ in 
Eq.~(\ref{eq:ap:Iq}) to find
\begin{equation}
  I_s \propto \lambda^{s/q-4}.
\end{equation}
So,
\begin{equation}
   \M20 = \M22 \propto \lambda^{2/q}.
\end{equation}
Since $\M21 = 0$,
\begin{equation}
   1-|V|^2 = \frac{\M20 B_u^2 + \M22 B_v^2}{\lambda^2},
\end{equation}
where $(B_u, B_v) = \lambda \vu$ is the projected baseline.  Therefore
\begin{equation}
   1-|V|^2 \propto \lambda^{2/q-2}. 
\end{equation}
If the disc is not face-on, the second-order moment along the minor axis is 
shrinked by a factor $\cos^2 i$, where $i$ is the inclination; but the
above demonstration still holds.

\section{Power-law accretion disc model}
\label{sec:pladm}

The FU~Ori disc model has been determined with an effective temperature $T(r)
= K r^{-q}$.  The influence of the central gap or of the star highly
depends on the constant $K$.  In the standard viscous disc model ($q = 3/4$) by
\citet{Shakura73}, 
\begin{equation}
   K = \sqrt[4]{\frac{3\mathcal{G}\Mstar\Mdot}{8\sigma\pi}},
\end{equation}
where $\Mstar$ stands for the mass of the star and $\Mdot$ for the
accretion rate.   If a deviation from $q = 3/4$ is observed, we should
carry out a proper modelling of the involved phenomena.  For the sake of 
simplicity, we assumed that $K$ is always determined by the viscous 
temperature at $1\,\AU$, so that
\begin{equation}
   T(r) = \frac{K}{(1\,\AU)^{-q}} \left(\frac{r}{1\,\AU}\right)^{-q}.
\end{equation}
We took
\begin{subequations}
\begin{align}
   \Mstar &= 1\,\Msun,\\
   \Mdot  &= 3\,\times\,10^{-5}\,\Msun/\yr.
\end{align}     
\end{subequations}

\bibliography{M3304.bib}

\begin{thebibliography}{20}
\expandafter\ifx\csname natexlab\endcsname\relax\def\natexlab#1{#1}\fi

\bibitem[{{Akeson} {et~al.}(2002){Akeson}, {Ciardi}, {van Belle}, \&
  {Creech-Eakman}}]{Akeson02}
{Akeson}, R.~L., {Ciardi}, D.~R., {van Belle}, G.~T., \& {Creech-Eakman}, M.~J.
  2002, \apj, 566, 1124

\bibitem[{{Akeson} {et~al.}(2000){Akeson}, {Ciardi}, {van Belle},
  {Creech-Eakman}, \& {Lada}}]{Akeson00}
{Akeson}, R.~L., {Ciardi}, D.~R., {van Belle}, G.~T., {Creech-Eakman}, M.~J.,
  \& {Lada}, E.~A. 2000, \apj, 543, 313

\bibitem[{{Armstrong} {et~al.}(1998){Armstrong}, {Mozurkewich}, {Rickard},
  {Hutter}, {Benson}, {Bowers}, {Elias}, {Hummel}, {Johnston}, {Buscher},
  {Clark}, {Ha}, {Ling}, {White}, \& {Simon}}]{Armstrong98}
{Armstrong}, J.~T., {Mozurkewich}, D., {Rickard}, L.~J., {et~al.} 1998, \apj,
  496, 550

\bibitem[{{Carleton} {et~al.}(1994){Carleton}, {Traub}, {Lacasse}, {Nisenson},
  {Pearlman}, {Reasenberg}, {Xu}, {Coldwell}, {Panasyuk}, {Benson},
  {Papaliolios}, {Predmore}, {Schloerb}, {Dyck}, \& {Gibson}}]{Carleton94}
{Carleton}, N.~P., {Traub}, W.~A., {Lacasse}, M.~G., {et~al.} 1994, in Proc.
  SPIE Vol. 2200, p. 152-165, Amplitude and Intensity Spatial Interferometry
  II, James B. Breckinridge; Ed., Vol. 2200, 152--165

\bibitem[{{Colavita}(2001)}]{Colavita01}
{Colavita}, M. 2001, American Astronomical Society Meeting, 198

\bibitem[{{Colavita} {et~al.}(1999){Colavita}, {Wallace}, {Hines}, {Gursel},
  {Malbet}, {Palmer}, {Pan}, {Shao}, {Yu}, {Boden}, {Dumont}, {Gubler},
  {Koresko}, {Kulkarni}, {Lane}, {Mobley}, \& {van Belle}}]{Colavita99}
{Colavita}, M.~M., {Wallace}, J.~K., {Hines}, B.~E., {et~al.} 1999, \apj, 510,
  505

\bibitem[{{Glindemann} {et~al.}(2000){Glindemann}, {Abuter}, {Carbognani},
  {Delplancke}, {Derie}, {Gennai}, {Gitton}, {Kervella}, {Koehler}, {Leveque},
  {Menardi}, {Michel}, {Paresce}, {Duc}, {Richichi}, {Schoeller}, {Tarenghi},
  {Wallander}, \& {Wilhelm}}]{Glindemann00}
{Glindemann}, A., {Abuter}, R., {Carbognani}, F., {et~al.} 2000, in Proc. SPIE
  Vol. 4006, p. 2-12, Interferometry in Optical Astronomy, Pierre J. Lena;
  Andreas Quirrenbach; Eds., Vol. 4006, 2--12

\bibitem[{{Hanbury Brown} {et~al.}(1974){Hanbury Brown}, {Davis}, \&
  {Allen}}]{Hanbury74}
{Hanbury Brown}, R., {Davis}, J., \& {Allen}, L.~R. 1974, \mnras, 167, 121

\bibitem[{{Haniff} {et~al.}(2000){Haniff}, {Baldwin}, {Boysen}, {George},
  {Buscher}, {Mackay}, {Pearson}, {Rogers}, {Warner}, {Wilson}, \&
  {Young}}]{Haniff00p}
{Haniff}, C.~A., {Baldwin}, J.~E., {Boysen}, R.~C., {et~al.} 2000, in Proc.
  SPIE Vol. 4006, p. 627-633, Interferometry in Optical Astronomy, Pierre J.
  Lena; Andreas Quirrenbach; Eds., Vol. 4006, 627--633

\bibitem[{{Labeyrie}(1975)}]{Labeyrie75}
{Labeyrie}, A. 1975, \apj, 196, 71

\bibitem[{{Lachaume} {et~al.}(2001){Lachaume}, {Malbet}, \&
  {Monin}}]{Lachaume02}
{Lachaume}, R., {Malbet}, F., \& {Monin}, J.-L. 2001, \aap, in press

\bibitem[{{Malbet} \& {Berger}(2002{\natexlab{a}})}]{Malbet02p}
{Malbet}, F. \& {Berger}, J.-P. 2002{\natexlab{a}}, in SPIE proceedings, Vol.
  {in press}

\bibitem[{{Malbet} \& {Berger}(2002{\natexlab{b}})}]{Malbet01p}
{Malbet}, F. \& {Berger}, J.-P. 2002{\natexlab{b}}, in {SF2A --- Scientific
  Highlights 2001}, ed. {F.~Combes, D.~Barret and F.~Thévenin} (EDP Sciences),
  457--460

\bibitem[{{Malbet} {et~al.}(1998){Malbet}, {Berger}, {Colavita}, {Koresko},
  {Beichman}, {Boden}, {Kulkarni}, {Lane}, {Mobley}, {Pan}, {Shao}, {van
  Belle}, \& {Wallace}}]{Malbet98}
{Malbet}, F., {Berger}, J.-P., {Colavita}, M.~M., {et~al.} 1998, \apj, 507, 149

\bibitem[{{Michelson}(1891)}]{Michelson91}
{Michelson}, A.~A. 1891, \pasp, 3, 274

\bibitem[{{Michelson}(1920)}]{Michelson20}
---. 1920, \apj, 51, 257

\bibitem[{{Michelson} \& {Pease}(1921)}]{Michelson21}
{Michelson}, A.~A. \& {Pease}, F.~G. 1921, \apj, 53, 249

\bibitem[{{Monnier} \& {Millan-Gabet}(2002)}]{Monnier02}
{Monnier}, J.~D. \& {Millan-Gabet}, R. 2002, \apj, 579, 694

\bibitem[{{Shakura} \& {Sunyaev}(1973)}]{Shakura73}
{Shakura}, N.~I. \& {Sunyaev}, R.~A. 1973, \aap, 24, 337

\bibitem[{{ten Brummelaar} {et~al.}(2000){ten Brummelaar}, {Bagnuolo},
  {McAlister}, {Ridgway}, {Sturmann}, {Sturmann}, \&
  {Turner}}]{tenBrummelaar00p}
{ten Brummelaar}, T.~A., {Bagnuolo}, W.~G., {McAlister}, H.~A., {et~al.} 2000,
  in Proc. SPIE Vol. 4006, p. 564-573, Interferometry in Optical Astronomy,
  Pierre J. Lena; Andreas Quirrenbach; Eds., Vol. 4006, 564--573

\end{thebibliography}

\end{document}